\documentclass[twocolumn,preprintnumbers,amsmath,amssymb,prl]{revtex4-1}
\usepackage{graphicx}

\usepackage{dcolumn}
\usepackage{bm}

\newcommand{\comment}[1]{}

\begin{document}

\title{Two-Element Mixture of Bose and Fermi Superfluids}

\author{Richard Roy}
\author{Alaina Green}
\author{Ryan Bowler}
\author{Subhadeep Gupta}

\affiliation{Department of Physics, University of Washington, Seattle, Washington 98195, USA}

\date{\today}

\begin{abstract}
We report on the production of a stable mixture of bosonic and fermionic superfluids composed of the elements $^{174}$Yb and $^6$Li which feature a strong mismatch in mass and distinct electronic properties. We demonstrate elastic coupling between the superfluids by observing the shift in dipole oscillation frequency of the bosonic component due to the presence of the fermions. The measured magnitude of the shift is consistent with a mean-field model and its direction determines the previously unknown sign of the interspecies scattering length to be positive. We also observe the exchange of angular momentum between the superfluids from the excitation of a scissors mode in the bosonic component through interspecies interactions. We explain this observation using an analytical model based on superfluid hydrodynamics.
\end{abstract}
\maketitle

Ultracold atomic gases offer excellent opportunities to produce and investigate quantum matter, allowing fundamental studies in many-body physics and quantum simulation \cite{bloc08}. Beginning with the cooling of atoms to Bose-Einstein condensation \cite{ande95,davi95} and later followed by the achievement of superfluidity in atomic Fermi gases \cite{zwer12}, such studies have found many parallels with the analogous superfluids of bosonic $^4$He and fermionic $^3$He as well as superconductors, familiar from condensed matter physics. While the goal of simultaneous superfluidity in mixtures of $^4$He-$^3$He still remains elusive due to strong inter-isotope interactions \cite{tuor02,ryst12}, Bose-Fermi superfluidity in an atomic gas isotopic mixture of $^7$Li-$^6$Li has recently been realized \cite{ferr14}. The original interest \cite{andr75,volo75} in such dual superfluid systems is now strongly intensified \cite{ozaw14,zhen14,zhan14,cui14,kinn15,cast15,tylu16}.

The extension of Bose-Fermi superfluidity to mixtures of different elements is experimentally challenging, but holds the potential to open a larger arena of scientific studies. A large mass ratio between the components is predicted to alter the interaction energy between the superfluids \cite{zhan14} as well as the character of excitations across the Bose-Einstein condensate to Bardeen-Cooper-Schreiffer (BEC-BCS) crossover for the fermion pairs \cite{cast15}. Specific interaction strengths and mass ratios can aid the detection of exotic states such as the Fulde-Ferrell-Larkin-Ovchinnikov (FFLO) phase \cite{ozaw14} and dark-bright solitons \cite{tylu16}. Furthermore, species-selective potentials for relative positioning and selective addressing make two-element systems more amenable for systematic studies.

In this paper, we report on the realization of a two-element Bose-Fermi superfluid mixture of $^{174}$Yb-$^6$Li. We measure their coupling by observing the interaction-induced frequency shift of the bosonic dipole mode. We also detect transfer of angular momentum between the superfluids through the excitation of a scissors mode in the bosonic component. The scattering length $a_F$ of the two-spin alkali $^6$Li fermionic system is tunable across the BEC-BCS crossover through a Feshbach resonance centered at $832\,$G, while the scattering length $a_B$ of the alkaline-earth-like bosonic Yb remains constant throughout. The combination of spin-half (Li,$\,^2S_{1/2}$) and spin-zero (Yb,$\,^1S_0$) electronic states allows external magnetic fields to be used as a convenient species-specific tool and results in a uniform interspecies scattering length $a_{BF}$ for all Li spin states.


The preparation of the dual superfluid involves significant extensions to our earlier cooling methods for the dual-species Yb-Li system \cite{hans13} (see supplemental material \cite{supp_bfsf16}). Briefly, we load laser-cooled atoms into a crossed-beam 1064$\,$nm optical dipole trap (ODT) optimized for efficient evaporative cooling by dynamically changing the trap shape \cite{roy16}. The relative polarizability and trap frequency of the bosonic and fermionic components are $\alpha_F/\alpha_B=2.2$ and $\omega_F/\omega_B=8$. We perform forced evaporative cooling of Yb and simultaneous sympathetic cooling of a single spin state of Li to quantum degeneracy at $B=330\,$G, achieving a mixture of up to $3 \times 10^5$ Yb atoms in a pure condensate and $2 \times 10^5$ Li atoms with $T/T_F \leq 0.2$.

At this stage, the Yb trap frequencies are $\left(\omega_x,\omega_y,\omega_z \right)_B = 2\pi \times \left(23,150,10 \right)\,$Hz, where gravity points in the -$y$ direction. For these parameters, the two species are thermally decoupled with cloud centers separated by $y_{0,F} - y_{0,B} = 11\, \mu$m due to the large differential gravitational sag stemming from the large mass mismatch.

To proceed towards pairing and condensation of the Fermi gas, we exploit the different electronic character of Yb and Li and individually address Li with external magnetic fields. First, we bring the system to $832\,$G and prepare a 50:50 spin mixture of Li in the two lowest hyperfine states, $|1\rangle$ and $|2\rangle$, using a radio-frequency (RF) pulse, and allow the resulting equal superposition state to decohere over 100$\,$ms. The magnetic insusceptibility of Yb ensures stability of the BEC at all magnetic fields as the trapping potential and $a_B=5.6\,$nm remain unchanged.

\begin{figure}
\begin{centering}
\includegraphics[width=1\columnwidth]{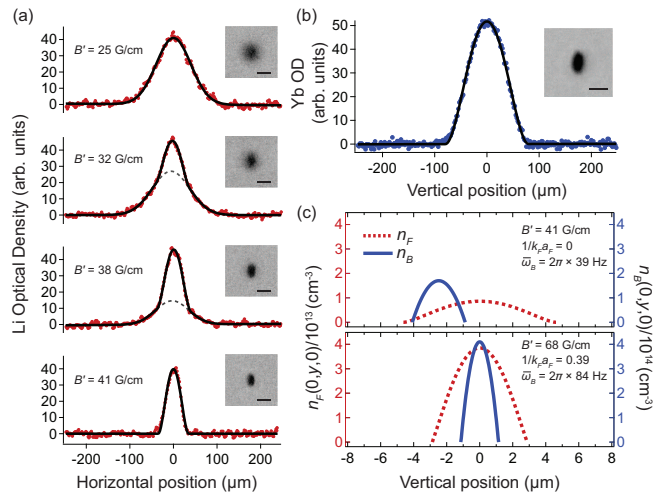}
\par\end{centering}
\caption{\label{fig:waterfall} (a) Gradient controlled forced evaporative cooling of a Li unitary Fermi gas.  We image the atoms at 690 G with 2 ms ToF (scale bar is 100$\,\mu$m). Solid lines are bimodal fits to doubly-integrated optical density (OD) profiles, with dashed lines indicating the thermal component. From top to bottom, the detected condensate fractions are $f_c = 0$, 0.2, 0.5, and 1. (b) Yb BEC of $1.1 \times 10^5$ (30 ms ToF) atoms coexisting with a Li superfluid of $0.8 \times 10^5$ atoms. Solid line is a fit to a pure Thomas-Fermi BEC profile. Gravity is in the vertical direction for all images in (a) and (b). (c) Calculated in-trap fermionic (dashed red lines) and bosonic (solid blue line) superfluid density profiles for (upper panel) the combined superfluid lifetime measurement at unitarity and (lower panel) the bosonic dipole oscillation measurements at 780\,G presented in Fig. \ref{fig:dipoleosc}. For the fermion we use the zero-temperature EoS $n_F(\mu_F,a_F)$ and the local density approximation to obtain $n_F(\vec{r})$ \cite{navo10,supp_bfsf16}, and for the boson a pure BEC profile in the Thomas-Fermi limit. The gradient of $B' = 41$ G/cm ensures complete overlap of the two clouds, while $B' = 68$ G/cm guarantees $y_{0,B}=y_{0,F}$.}
\end{figure}

Next we perform independent forced evaporative cooling of Li at $832\,$G using a magnetic field gradient $B'$ in the vertical direction \cite{hung08}. This gradient reduces the trap depth for Li and moves the cloud center towards the Yb BEC. Typically, we ramp $B'$ from zero to its final value over $500\,$ms, and hold for $200\,$ms. To perform thermometry of Li, we ramp the magnetic field in trap from $832\,$G to $690\,$G, where $1/k_Fa_F = 2.9$, in $5\,$ms, and image the resulting molecular cloud in time-of-flight (ToF). Here $k_F = \sqrt{2m_F E_F}/\hbar$ is the Fermi wavevector and $k_B T_F = E_F = \hbar \bar{\omega}_F (3N_F)^{1/3}$ the Fermi energy of a harmonically-trapped, spin-balanced Fermi gas with $N_F$ atoms and geometric mean trap frequency $\bar{\omega}_F$. Fig. \ref{fig:waterfall}(a) displays a progression for different final $B'$ values towards the detection of a pure molecular BEC (mBEC) consisting of $0.4 \times 10^5$ molecules, coexisting with a pure Yb superfluid of $1.1 \times 10^5$ atoms (Fig. \ref{fig:waterfall}(b)) with an applied gradient of $B' = 41\,$G/cm ensuring complete interspecies overlap.

We infer superfluidity of the Fermi gas at unitarity by comparing the observed entropy of the mBEC with the equation of state (EoS) of the unitary Fermi gas \cite{ku12,haus08}. Importantly, we observe no difference in molecular condensate fraction $f_c$ if we remove the gradient $B'$ before ramping from 832 G to 690 G. For $B'= 41$ G/cm, we estimate that removing the gradient increases the trap depth for Li by a factor of 6. Consequently, we conclude that no evaporative cooling occurs during the ramp to the weakly interacting BEC regime for the measurements in Fig. \ref{fig:waterfall}(a). Therefore, the observed entropy of the mBEC gives an upper bound on the entropy of the initial unitary Fermi gas, since the ramp is at best adiabatic.

To determine the entropy of the molecular condensate at 690 G, we calculate the relevant quantities in the Thomas-Fermi limit. This is justified since $n_m(0)a_m^3 = 0.001$, where $n_m(0)$ is the peak mBEC density and $a_m = 0.6 a_F$ is the molecule-molecule scattering length \cite{petr04}. The thermal fraction of the coldest Li clouds is below our detection limit of $1-f_c = 0.15$, implying that the total entropy in the trap, including the effects of interactions \cite{carr04,supp_bfsf16}, is $S/(N_Fk_B) \leq 0.55$, which is well below the critical entropy for a unitary Fermi gas at the superfluid transition in a harmonic trap $S_c/(N_Fk_B) = 1.70$ \cite{ku12}. The measurements in \cite{ku12} determine the EoS $S(T/T_F)$ for trap-averaged reduced temperatures above $T/T_F = 0.15$, at which point $S/(N_Fk_B) = 0.92$, and are thus not applicable at our measured entropy. If instead we compare with the EoS calculated in \cite{haus08}, which agrees well with \cite{ku12}, we determine an upper bound on the temperature at unitarity of $T \leq 0.12 T_F = 0.55 T_{c,F}$, where $T_{c,F} = 0.22 T_F$ is the critical temperature for superfluidity at unitarity in a harmonic trap \cite{supp_bfsf16}.

To measure the stability of the dual superfluid at unitarity, we first adiabatically increase the ODT power by 50\% to inhibit Yb evaporation and subsequently hold the overlapped clouds in the trap for a variable time at $832\,$G before ramping to $690\,$G and imaging in ToF. The upper panel in Fig. \ref{fig:waterfall}(c) depicts the relative Bose and Fermi superfluid density profiles in the vertical dimension for this situation. From exponential fits to the observed condensate number evolution we determine lifetimes of $1.8\,$s and $0.7\,$s for Yb and Li, respectively.

The Yb-Li dual superfluid system features a small BEC inside a larger Fermi superfluid. For the lifetime measurements at unitarity discussed above, the relative  in-trap cloud radii are $R_B/R_F = 0.36$, where $R_{\beta} = (2\mu_{\beta}/m_\beta\bar{\omega}_\beta^2)^{1/2}$ for $\beta = B$ and $F$, $\mu_F = \sqrt{\xi} E_F$ is the chemical potential for a zero-temperature unitary Fermi gas, $\xi = 0.37$ is the Bertsch parameter \cite{ku12,zurn13}, and $\mu_B$ is the chemical potential for the Yb BEC. Additionally, the effects of interspecies mean-field interactions are small, with $V_{BF(FB)}(0)/\mu_{F(B)} = 0.07(0.14)$, where $V_{BF(FB)}(0)$ is the peak interspecies mean field interaction energy of Yb on Li (Li on Yb).

To probe elastic interactions in the dual superfluid we selectively excite vertical center-of-mass (dipole) oscillations in the bosonic component and measure the oscillation frequency with and without the presence of the overlapped fermionic component \cite{ferr14}. Due to the large ratio of trap frequencies for Li and Yb, we can achieve this species-selective excitation by increasing the trapping laser power $P$ on a timescale that is diabatic (adiabatic) for Yb (Li), where the diabatic excitation arises from the power-dependent gravitational sag, $y_{0,B} \propto 1/P$. Prior to the Yb excitation, we increase the magnetic gradient to the value $B' = 68$ G/cm that ensures $y_{0,F}=y_{0,B}$ (Fig.\,\ref{fig:waterfall}(c), lower panel), concurrent with an adiabatic increase of $P$ to prevent spilling of Li due to the strong gradient.

\begin{figure}
\begin{centering}
\includegraphics[width=1\columnwidth]{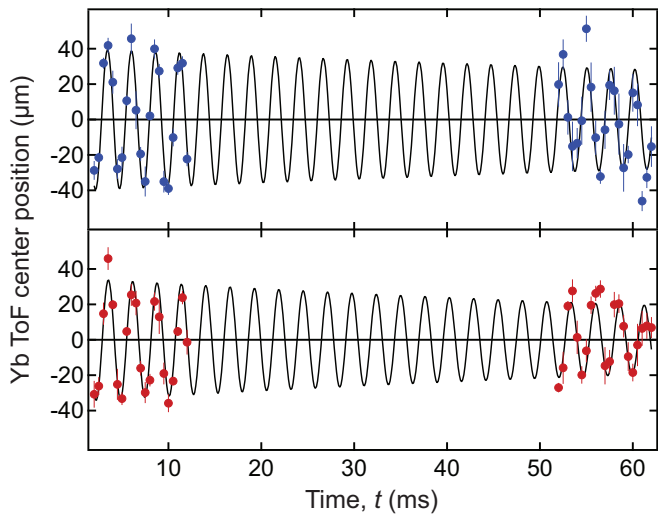}
\par\end{centering}
\caption{\label{fig:dipoleosc} Yb BEC dipole oscillations without (blue circles, upper panel) and with (red circles, lower panel) the Li superfluid at $780\,$G. Center-of-mass position measured at 30 ms ToF (solid lines are fits, see text).}
\end{figure}

Fig. \ref{fig:dipoleosc} shows dipole oscillation measurements of the Yb BEC with and without the Fermi superfluid at $780\,$G, where $1/k_F a_F = 0.39$. We perform these measurements at 780$\,$G instead of 832$\,$G in order to enhance the effect of Li on the Yb oscillations \cite{ferr14}. For these measurements, after the species-selective excitation, the superfluids are held in a trap with $\left(\omega_x,\omega_y,\omega_z \right)_B = 2\pi \times \left(59,388,26 \right)\,$Hz for a variable time $t$ before ToF imaging \cite{foot3_bfds}. During the $30\,$ms ToF for Yb, the in-trap momentum oscillation converts into an oscillation of the vertical position. For the measurements without Li, we keep the experimental preparation of the dual superfluid the same, and remove the fermionic component by spilling with a strong gradient before initiating the dipole oscillation.

We fit exponentially-damped sinusoids to the measurements in Fig. \ref{fig:dipoleosc} and extract $\omega_{y,B}/2\pi = 387.7(3)\,$Hz (without Li) and $\omega_{y,B}'/2\pi = 381.3(4)\,$Hz (with Li), yielding a reduction of the dipole oscillation frequency of $1.7(2)\,$\%. From the amplitude of the oscillation in ToF, we determine the maximum in-trap velocity and displacement to be $1.2\,$mm/s and $0.5\,\mu$m for the Yb BEC with a vertical Thomas-Fermi radius $R_{B,y} = 1.1\,\mu$m. The small amplitude of oscillation ensures that the Bose probe remains well localized inside the larger Fermi cloud (Fig.\,1(c), lower panel).

From the decay constant $\tau$ for the oscillations with Li, we find that $\omega_{y,B}'\tau = 250$, which ensures that our determination of the dipole frequency is unaffected by the decay. The finite quality factor is most likely due to anharmonicities of the ODT potential, as the damping times with and without Li are within error of each other. While the maximum Yb velocity is an order of magnitude below the critical velocity of this Bose-Fermi superfluid system \cite{cast15,dele15}, we cannot rule out the possibility of dissipation due to a finite Li thermal component.

To model the effect of Li on the Yb dipole oscillations, we adopt a mean-field treatment, in which Yb experiences an effective potential $V_B(\vec{r}) = V_{T,B}(\vec{r}) + g_{BF}n_F(\vec{r})$, where $V_{T,B}(\vec{r})$ is the Yb optical potential, $g_{BF} = 2\pi \hbar^2 a_{BF} / m_{BF}$, $m_{BF}$ is the Yb-Li reduced mass, and $n_F(\vec{r})$ is the fermion density. In the local density approximation, the spatial curvature of the Fermi superfluid density gives rise to a shift of the bosonic dipole oscillation frequency given by \cite{ferr14}
\begin{align}
\frac{\delta\omega_{y,B}}{\omega_{y,B}} = - \frac{g_{BF}}{2} \frac{\alpha_F}{\alpha_B} \left. \frac{dn_F}{d\mu_F} \right\rvert_{\mu_F(0)}. \label{eqn:freqshift}
\end{align}
Here $\delta\omega_{y,B} = \omega_{y,B}'-\omega_{y,B}$, $\mu_F(\vec{r})=\mu_F(0)-V_{T,F}(\vec{r})$ is the local chemical potential, and $V_{T,F}(\vec{r}) = (\alpha_F/\alpha_B)V_{T,B}(\vec{r})$ is the Li optical potential. Using the magnitude of $a_{BF}$ derived from interspecies thermalization measurements \cite{ivan11,hara11}, $|a_{BF}| = 15(2)\,a_0$ \cite{foot2_bfds}, this model predicts a frequency shift at $1/k_Fa_F = 0.39$ of absolute value $2.0(3)\,$\%. The uncertainty in the prediction comes entirely from the uncertainty in the measurement of $|a_{BF}|$. The predicted magnitude is in good agreement with the measurement in Fig.\,\,\ref{fig:dipoleosc}, and the direction of the observed frequency shift determines the previously unknown sign of the Yb-Li $s$-wave scattering length to be positive.


The mean field interaction exerted on the fermions by the oscillating BEC also modulates the potential felt by the fermions. However, due to the large value of $\omega_F/\omega_B$, this modulation can be adiabatically followed by Li. Consistent with this picture, we observe no motion from backaction on Li.

\begin{figure}
\begin{centering}
\includegraphics[width=1\columnwidth]{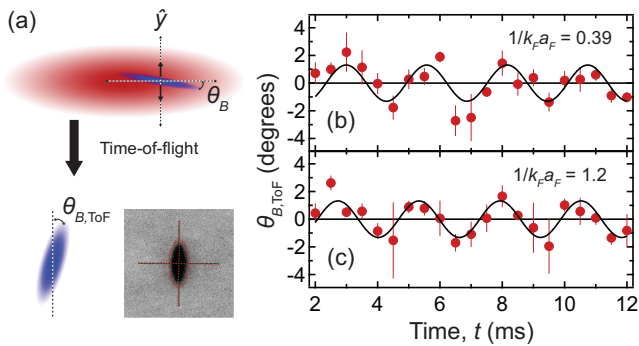}
\par\end{centering}
\caption{\label{fig:scissors} (a) While executing dipole oscillations along $\hat{y}$, the tilt angle of the Yb BEC $\theta_B$ exhibits a modulation due to interactions with the horizontally-offset Li superfluid. This in-trap angle modulation maps proportionally to a tilt $\theta_{B,\text{ToF}}$ for long ToF. In the absorption image, the red axes are aligned to the major and minor axes of the BEC tilted 1.3$^\circ$ clockwise with respect to the trap eigenaxes (black). (b)-(c) Observation of a scissors mode excitation in Yb due to interaction with Li at $780\,$G ($1/k_Fa_F = 0.39$) and $720\,$G ($1/k_Fa_F = 1.2$).}
\end{figure}

The shift in the dipole oscillation frequency clearly demonstrates the coupling between the two superfluids. Additionally, we observe a modulation of the angular orientation of the Yb BEC due to interactions with Li. Our observations are shown in Fig.\,\,\ref{fig:scissors}(b), where we plot the tilt angle $\theta_{B,\text{ToF}}$ of the Yb BEC at $30\,$ms ToF for the same dataset used in Fig.\,\,\ref{fig:dipoleosc}. The presence of Li leads to a modulation of this angle during the dipole oscillation. By fitting to a pure sine wave, we extract tilt angle modulation frequencies of $\omega_\theta/\omega_{y,B}' = 1.02(3)$ with modulation amplitude of $1.3(3)$ degrees. The amplitude of a fit to the data without Li is consistent with zero modulation. Similar behavior is observed at $720\,$G (Fig.\,\,\ref{fig:scissors}(c)) where $1/k_Fa_F = 1.2$, with the fit returning a modulation frequency of $\omega_\theta/\omega_{y,B}' = 1.01(3)$ and amplitude of $1.3(3)$ degrees. For each field, we present the frequency measurement as a ratio with respect to the mean-field-shifted frequency $\omega_{y,B}'$ at that field. Because the $\omega_\theta$ measurement precision is comparable with the frequency shift $\delta\omega_{y,B}$, the result is consistent with both $\omega_{y,B}$ and $\omega_{y,B}'$.

The small elastic scattering cross-section between the two species allows us to rule out collisional hydrodynamics for this observation and instead favor superfluid hydrodynamics as the cause. We interpret the tilt angle modulation as the excitation of a scissors mode of oscillation in the bosonic component driven by interaction with the fermionic component. In trapped atomic systems, the scissors mode is a small amplitude angular oscillation about a principal trap axis, and has previously been utilized to demonstrate and study superfluidity in ultracold atomic gases \cite{guer99,mara00,modu02,wrig07}. While the resulting flow is irrotational, the superfluid acquires angular momentum that oscillates at the same frequency as the in-trap angle $\theta_B(t)$ \cite{supp_bfsf16}.

Our observations imply the existence of a horizontal offset between the two cloud centers which provides a finite impact parameter and therefore non-zero torque during the Yb dipole oscillation (Fig.\,\,\ref{fig:scissors}(a)). In our system, the horizontal cloud offset arises due to a combination of slight angular offsets of both the magnetic field gradient direction and the nearly vertical principal axis of the ODT with respect to the direction of gravity.  Since our experimental geometry forces a near degeneracy between the dipole oscillation frequency $\omega_{y,B}$ and that of the scissors mode $\omega_s$, the sinusoidal motion of the Yb center-of-mass within an offset Li cloud can resonantly drive the scissors mode.

In order to explain the interspecies-interaction-driven scissors mode, we extend the analytical model based on superfluid hydrodynamics developed in \cite{guer99} to include the interaction with the fermionic superfluid \cite{supp_bfsf16}. The resulting linear-response dynamics for the BEC in-trap tilt angle, $\theta_B$, are identical to those of a sinusoidally-driven harmonic oscillator,
\begin{align}
\frac{d^2\theta_B}{dt^2} = -\omega_s^2\theta_B + g(x_0,y_0)\omega_{x,B}^2 \cos(\omega_{y,B}'t) \label{eqn:scissorsdynamics1},
\end{align}
where $\omega_s = (\omega_{y,B}^2+f(x_0)\omega_{x,B}^2)^{1/2}$, $x_0$ is the fixed horizontal cloud displacement, and $y_0$ is the in-trap amplitude of the vertical dipole oscillation. The functions $f$ and $g$ encompass the interspecies interaction effects over the BEC density distribution. Since $|f|$ is at most of order unity \cite{supp_bfsf16}, $\omega_s \approx \omega_{y,B}$, which clearly reveals the near-resonant nature of the sinusoidal drive.


In this treatment we neglect backaction onto the Li cloud since the Yb angular oscillation frequency is far detuned from the corresponding scissors mode in Li. This implies that the angular momentum imparted to Li from the interaction with Yb is adiabatically transferred to the trap. This effect and the dynamics in Eqn. (\ref{eqn:scissorsdynamics1}) are reproduced in a full numerical simulation of the coupled superfluid dynamics \cite{foot_forbes16}.

The observed scissors mode amplitude in Fig. \ref{fig:scissors} is consistent with a response at the driving frequency $\omega_{y,B}'$, but does not display the linear growth of amplitude in time as one would expect from equation (\ref{eqn:scissorsdynamics1}). We interpret this to be indicative of damping in the scissors mode. Thus, in order to apply our model to the observations, we add a heuristic damping term $-\omega_s\dot{\theta}_B/Q_s$ to the right hand side of equation (\ref{eqn:scissorsdynamics1}), where $Q_s$ is the quality factor of the scissors mode. We then find that our model reproduces the amplitude observed in Fig.\,\,\ref{fig:scissors}(b) for a displacement of roughly twice the Yb Thomas-Fermi radius and $Q_s = 4$ \cite{supp_bfsf16}. Furthermore, for this displacement our model predicts the driving force at 720 G to be only 10-20\% larger than that at 780 G, which is consistent with Fig.\,\,\ref{fig:scissors}(c).



We have analyzed the consequences of a horizontal offset between the cloud centers for the frequency shift measurements presented in Fig. \ref{fig:dipoleosc} \cite{supp_bfsf16}. On the BEC side of the Feshbach resonance, the mean-field frequency shift is remarkably robust to horizontal cloud displacement, falling to a minimum at large displacements of $65\%$ of the value predicted by Eqn. (\ref{eqn:freqshift}) for $1/k_Fa_F = 0.39$.  Our dipole oscillation measurements are therefore consistent with the simultaneous observation of the driven scissors mode.


In conclusion, we have established a stable two-element Bose-Fermi superfluid system of $^{174}$Yb-$^6$Li and studied the frequency shift of dipole oscillations and the excitation of a scissors mode due to interspecies interactions. Extension of these methods can allow investigation of higher order excitations and sound propagation in the dual superfluid \cite{volo75}. Our experiments also highlight the use of the differences in mass and electronic structure for the selective excitation and controlled spatial overlap of the components, opening new perspectives for investigating the phase diagram of the Bose-Fermi superfluid system \cite{kukl03,rama11,tylu16}.

\begin{acknowledgements}
We thank Martin Zwierlein and Michael McNeil Forbes for useful discussions and critical reading of the manuscript, and the authors of \cite{ku12,haus08} for providing us with their data. This work was supported by grants from the NSF, AFOSR, and ARO-MURI.
\end{acknowledgements}



\noindent {\it Note:} Recently, we became aware of the realization of a two species mixture of Bose-Fermi superfluids in a $^{41}$K-$^6$Li system \cite{yao16}.
%

\pagebreak
\onecolumngrid
\clearpage
\begin{center}
\large \textbf{Supplemental Material for: \\ Two-Element Mixture of Bose and Fermi Superfluids}
\end{center}
\vspace{.3in}
\twocolumngrid

\renewcommand*{\citenumfont}[1]{S#1}
\renewcommand*{\bibnumfmt}[1]{[S#1]}

\section{Experimental Details}
As noted in the main text, several significant upgrades were made to our previous Yb-Li dual species system \cite{hans13s} in order to perform the current work. Recently \cite{roy16s} we implemented a new crossed optical dipole trap (ODT) geometry, leveraging the time-averaged, or ``painted," optical potential of a rapidly moving laser beam in the horizontal plane to achieve highly efficient Yb quantum degenerate gas production. Briefly, we apply a frequency modulated (FM) waveform to the input of a voltage controlled oscillator in order to create a time-dependent frequency for driving the acousto-optic modulator (AOM), resulting in a time-dependent deflection angle of the first-order diffracted beam. The FM waveform used results in a nearly perfect parabolic time-averaged trapping potential in the painting dimension. In order to extend this technique to Li, we increase the painting modulation frequency to 60 kHz from 10 kHz, as the trap frequency of Li is 8 times that for Yb. 

Additionally, we implement sub-Doppler laser cooling of Li using a gray molasses operating on the $^2S_{1/2}\,\rightarrow\,^2P_{1/2}$ transition \cite{grie13s,burc14s,siev15s}. After a magneto-optical trap (MOT) loading time of 2$\,$s and application of the gray molasses for 300$\,\mu$s, we have $N_F = 2-3 \times 10^8$ Li atoms at a temperature of $T = 60-70\,\mu$K before loading into the ODT. After a 100$\,$ms hold at 330$\,$G in the ODT at the full power of $P=55\,$W per beam and a painting amplitude of $h = 13 w_0 = 390\,\mu$m, where $w_0 = 30\,\mu$m is the Gaussian beam waist, we have $N_F = 5 \times 10^6$ Li atoms at $T = 60\,\mu$K in a 50:50 mixture of the two lowest hyperfine states $|1\rangle$ and $|2\rangle$. We then perform forced evaporative cooling of Li at 330$\,$G by reducing the ODT power a factor of 10 in 1$\,$s, after which $N_F = 6 \times 10^5$ atoms remain. The bias field and field gradient in our experiment point in the vertical direction.

Keeping the ODT power at a low value, we subsequently load the Yb MOT for 5$\,$s, during which the Li number in the ODT is unchanged as collisions are absent at zero field in the $|1\rangle$-$|2\rangle$ mixture. In 300$\,$ms before the Yb compression and ODT loading phase, we adiabatically increase the ODT depth to its maximum value, maintaining the painting amplitude of 390$\,\mu$m. After loading $20-30 \times 10^6$ Yb atoms into the ODT, we immediately blast away the state $|1\rangle$ Li atoms, for two reasons. First, since we sympathetically cool Li with Yb, we do not want to end the Yb evaporative cooling sequence with a larger number of Li than Yb. Second, in order to enter the strongly interacting regime for Li, we must sweep the magnetic field up to $832\,$G to access the $|1\rangle$-$|2\rangle$ Feshbach resonance. With a mixture of Yb and states $|1\rangle$ and $|2\rangle$ of Li, this sweep results in large inelastic loss as there is a strong enhancement of 3-body collisions on the molecular, or $a_F > 0$, side of the $832\,$G Feshbach resonance.

We perform forced evaporative cooling of Yb in a dynamically shaped optical potential by simultaneously reducing the ODT power in an exponential fashion while linearly decreasing the amplitude of painting \cite{roy16s}, resulting in very efficient cooling of both Yb and Li. Near the point where Yb becomes quantum degenerate, we finish evaporative cooling by increasing the painting amplitude considerably from its minimum value of 80$\,\mu$m to 450$\,\mu$m while only slightly increasing the ODT power from 1$\,$W to 1.5$\,$W. At this point, the gravitational sag of Yb is such that the two clouds are spatially separated and thermally decoupled.

\begin{figure}
\centering
\includegraphics[width=1\columnwidth]{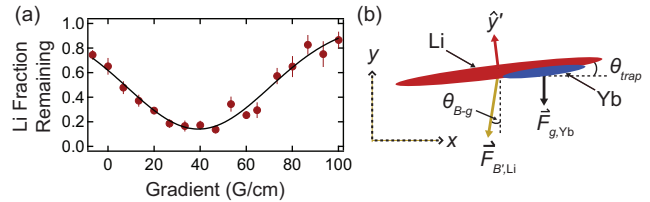}
\caption{\label{fig:gradcal} Calibration of the interspecies overlap with a magnetic field gradient. (a) Fraction of Li atoms remaining after a 500 ms hold at 740 G with Yb versus gradient $B'$. The solid line is a Gaussian fit to the data. The minimum value is attained for a gradient $B' = 39$ G/cm that is lower than the theoretical gradient for perfect overlap $B' = 68$ G/cm, due to small angular misalignments in our experiment. (b) Depiction of relative angles between the bias field direction, gravity, and the nearly-vertical trap eigenaxis for $B'=68$ G/cm. If $\theta_{B-g} \neq 0$, the two clouds experience a different projection of the respective vertical force onto the horizontal trap axes, while the vertical positions of Li and Yb are the same. We estimate these angular misalignments in our apparatus to be a few degrees.}
\end{figure}

To prepare the 50:50 mixture of states $|1\rangle$ and $|2\rangle$ of Li at this point, we ramp the magnetic field to $832\,$G and perform a 10$\,$ms Landau-Zener RF frequency sweep centered at 76.271$\,$MHz, resulting in the superposition state $(|1\rangle+|2\rangle)/\sqrt{2}$. We measure the decoherence of this superposition by measuring the aspect ratio of the Li cloud in ToF using absorption imaging at 832$\,$G as a function of time after the RF pulse. Since the single-state Li cloud begins with $T \leq 0.2T_F$, the cloud immediately becomes hydrodynamic upon decohering into a spin mixture, resulting in an inversion of the aspect ratio with respect to a non-interacting Fermi gas. By this method, we find that the entire cloud is decohered after 100$\,$ms.

To investigate the gradient-tunable interspecies overlap, we measure the inelastic coupling between the Yb and Li clouds at 740 G \cite{khra12s} as a function of $B'$ (Fig. \ref{fig:gradcal}(a)). Specifically, we hold non-degenerate clouds of Li and Yb together for 500 ms at 740 G with variable applied gradients $B'$ in the same trap used for the dipole oscillation measurements in Fig. 2 of the main text. At this field, the Li atom loss is mostly due to atom-molecule Yb-Li$_2$ inelastic collisions. The broadness of the dip in Li atom number shown in Fig. \ref{fig:gradcal}(a) reflects both the two cloud radii and the nonlinear nature of two-body inelastic loss. Importantly, the minimum does not occur for the gradient at which one would expect to see the two clouds perfectly overlapped, $B_\text{opt}' = (m_B g/\mu_B)(\alpha_F/\alpha_B + m_F/m_B) = 68$ G/cm, where $\mu_B$ is the Bohr magneton. This is due to the fact that the forces of gravity (Yb) and the gradient (Li) are not perfectly parallel to each other, or to the nearly-vertical trap eigenaxis, resulting in a different projection of the respective forces onto the horizontal trap axes (see Fig. \ref{fig:gradcal}(b)).

 Additionally, when we perform the same dipole oscillation measurement as in Fig.\,2 of the main text without $B'$, we find no difference between the Yb oscillation frequencies with and without Li, thus further confirming the necessity of the magnetic field gradient for establishing coupling between the superfluids.

\section{L\lowercase{i} Thermometry}

As described in the main text, after we apply the gradient $B'$ to evaporatively cool Li, our method of thermometry involves performing a magnetic field ramp from unitarity (832 G) to the weakly interacting BEC regime (690 G). This ramp is very close to linear and takes a time $\tau_r = 5$ ms. For the lowest Li trap frequency $\omega_{z,F} = 2\pi \times 80$ Hz, this gives $ \omega_{z,F} \tau_r = 3$, while for the tightest trapping direction (vertical) $\omega_{y,F}\tau_r = 38$. In order to check experimentally that the ramp does not greatly affect the cloud properties, we try various ramp times from $\tau_r = 5$ to $15$ ms, and find no discernible difference in the detected condensate fractions. For ramp times longer than 15 ms, Yb-Li$_2$ atom-molecule and Li$_2$-Li$_2$ molecule-molecule inelastic collisions become problematic.

\begin{figure}
\begin{centering}
\includegraphics[width=1\columnwidth]{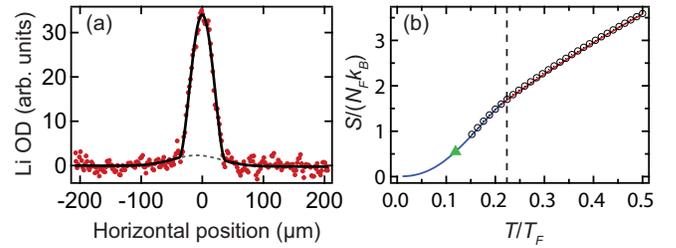}
\par\end{centering}
\caption{\label{fig:thermometry} (a) Doubly-integrated optical density (OD) profile of a molecular BEC with a condensate fraction of $f_c = 0.85$, taken with the same trap parameters as those in Fig. 1(a) of the main text, and at 2 ms ToF. (b) Per-particle, trap-averaged entropy $S/(N_Fk_B)$ as a function of $T/T_F$ for the unitary Fermi gas. (Open black circles) Measured EoS from \cite{ku12s}, transformed to trap-averaged quantities. (Black dashed line) The measured superfluid transition temperature at unitarity, $T_{c,F}/T_F = 0.22$. (Solid line) Calculated EoS from \cite{haus08s}. The blue(red) section corresponds to the superfluid(normal) phase, while the purple represents a multivalued region near the superfluid phase transition. (Green triangle) The upper bound on our coldest unitary Fermi gases $S/(N_F k_B) \leq 0.55$, indicating $T \leq 0.12 T_F = 0.55T_{c,F}$.}
\end{figure}

In order to determine an upper bound for the entropy of our purest molecular BECs, we need to know how small of a thermal fraction we can detect. Fig. \ref{fig:thermometry}(a) shows a typical doubly-integrated optical density (OD) trace in which we determine a thermal fraction of $1-f_c=15\%$. Next, we check that we can treat the mBEC in the Thomas-Fermi limit, using the parameters for the pure mBEC in Fig. 1(a) of the main text. The peak density in the condensate is $n_m(0) = \mu_m/g_m$, where $\mu_m = \hbar \bar{\omega}_F (15N_0 a_m/\bar{a}_\text{HO})^{2/5}/2$ is the chemical potential of the mBEC with $N_0 = 0.4 \times 10^5$ condensed molecules, $a_m = 0.6a_F = 0.6 \times 1420\,a_0$ at 690 G, $\bar{a}_\text{HO} = \sqrt{\hbar/(2m_F\bar{\omega}_F)}$ is the (geometric) mean harmonic oscillator length, and $g_m = 4\pi \hbar^2 a_m/(2m_F)$. Using $\bar{\omega}_F = 8\bar{\omega}_B = 2\pi \times 260$ Hz, we find that $\mu_m/k_B = 290$ nK, $n_m(0) = 1.3 \times 10^{13}$ cm$^{-3}$, and $n_m(0)a_m^3 = 0.001$. Thus, the detected mBEC is indeed within the Thomas-Fermi limit.

Even though the interaction parameter $n_m(0)a_m^3$ is small, interactions can still greatly affect the entropy, since it is the ratio of $\mu_m$ to $k_BT$ that determines whether the gas is mostly in the free-particle regime, the phonon regime, or in between \cite{dalf99s}. In order to assess this ratio for our measurements, we estimate the temperature $T$ of the mBEC in Fig. 1(a) of the main paper using $T = (1-f_c)^{1/3} T_{c,m} = 230$ nK, where $T_{c,m} = 0.94 \hbar \bar{\omega}_F (N_m)^{1/3}/k_B$ is the critical temperature for Bose-Einstein condensation of the molecules, $N_m = N_0/f_c$ is the total number of molecules, and we use our minimum detected thermal fraction of $15\%$. The entropy of the molecular BEC, including mean-field interactions, is \cite{carr04s}
\begin{align}
S = N_mk_B (1-f_c)\left(\frac{4\zeta(4)}{\zeta(3)} + \frac{3\mu_m}{k_BT} \right),
\end{align}
which is valid for $\mu_m/(k_BT) < 10$. Using $N_m = N_F/2$, this procedure gives us an upper bound $S/(N_F k_B) \leq 0.55$ for the entropy per particle at unitarity.

In order to compare this upper bound for the entropy with the measured equation of state at unitarity \cite{ku12s}, the measured homogeneous quantities must be converted to the trap-averaged ones. To do so, we use an interpolating function in between the data points
from \cite{ku12s}, and use the known virial expansion for values of $\beta \mu < -1.6$, where $\beta = 1/k_BT$. From the measured critical value $(\beta \mu)_c^{-1} = 0.40$ at the superfluid transition, we find that $T_{c,F}/T_F = 0.22$ and $S_c/(N_F k_B) = 1.70$, where $k_BT_F = \hbar \bar{\omega}_F (3N_F)^{1/3}$, in agreement with the values given in \cite{ku12s,kuthesiss}. The resulting EoS $S(T/T_F)$ is shown in Fig. \ref{fig:thermometry}(b). Since the experimentally determined EoS does not exist for temperatures lower than $0.15T_F$, we use the calculated EoS from \cite{haus08s}, also shown in Fig. \ref{fig:thermometry}(b), to ascribe a value of $T/T_F$ to our upper bound on $S/(N_Fk_B)$. As discussed in \cite{ku12s}, the calculations in \cite{haus08s} agree quite well with the measured EoS. From this we arrive at the upper bound on the temperature of the unitary Fermi gas of $T \leq 0.12 T_F = 0.55T_{c,F}$.

\section{Relevant scattering lengths}
\begin{figure}
\begin{centering}
\includegraphics[width=1\columnwidth]{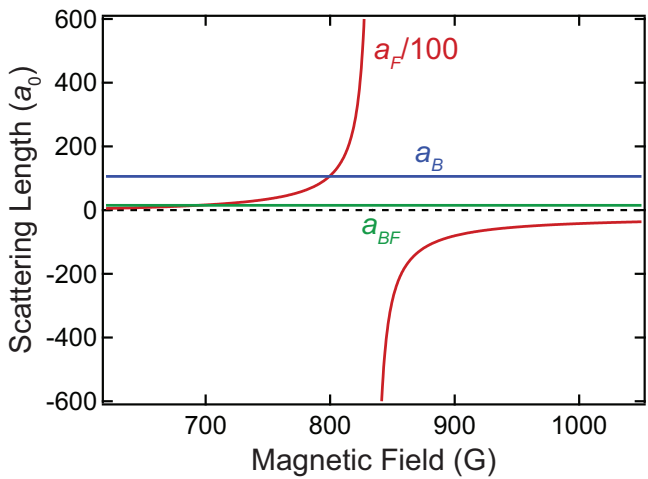}
\par\end{centering}
\caption{\label{fig:scatlengths} Relevant scattering lengths for the $^{174}$Yb-$^6$Li dual superfluid system in the vicinity of the 832$\,$G Feshbach resonance in $^6$Li. The magnetic field independent scattering lengths are $a_B = 106\,a_0$ for Yb-Yb (blue) and $a_{BF} = 13\,a_0$ for Yb-Li (green). Note that the Li-Li scattering $a_F(B)$ is scaled down by a factor of 100 in the plot.}
\end{figure}

Fig. \ref{fig:scatlengths} displays the Bose-Bose ($a_{B}$), Bose-Fermi ($a_{BF}$), and Fermi-Fermi ($a_F$) $s$-wave scattering lengths across the BEC-BCS crossover in Li.

\section{\label{sec:LiDensProf} L\lowercase{i} density profiles for arbitrary $\bm{1/k_Fa_F}$}

To get the shape of the Fermi superfluid for arbitrary $1/k_Fa_F$, where $k_F = \sqrt{2m_F\bar{\omega}_F(3N_F)^{1/3}/\hbar}$, we use the zero-temperature equation of state from \cite{navo10s}, parametrized by writing the pressure of a spin-unpolarized, two-component Fermi gas as $P(\mu_F,a_F) = 2P_0(\mu_F)h(\delta)$, where $\mu_F$ is the fermionic chemical potential excluding the contribution from the molecular binding energy on the BEC side, $P_0(\mu_F)$ is the single-component, ideal Fermi gas pressure, and $\delta = \hbar/(\sqrt{2m_F\mu_F}a_F)$. Using the Gibbs-Duhem relation, $n_Fd\mu_F = dP$, one can derive the spatial profile for a trapped, stongly-interacting Fermi gas within the local density approximation,
\begin{align}
\frac{n_F(\vec{r})}{n_0(0)} = (\delta(\vec{r})k_Fa_F)^{-3}\left( h(\delta(\vec{r})) - \frac{\delta(\vec{r})}{5} h'(\delta(\vec{r})) \right), \label{eqn:zerotempprofile}
\end{align}
where $n_0(0)$ is the peak density for a harmonically-trapped, two-component ideal Fermi gas,
\begin{align}
\delta(\vec{r}) = \delta_0\left(1-(\delta_0k_Fa_F)^2\sum\limits_{i} \left(\frac{x_i}{R_{0,F,i}}\right)^2 \right)^{-1/2},
\end{align}
$\delta_0 = \hbar/(\sqrt{2m_F\mu_F(0)}a_F)$, $\mu_F(\vec{r}) = \mu_F(0)-V_{T,F}(\vec{r})$, and $R_{0,F,i} = \sqrt{2E_F/(m_F\omega_{i,F}^2)}$ is the Fermi radius of the ideal Fermi gas with Fermi energy $E_F = \hbar \bar{\omega}_F (3N_F)^{1/3}$. Lastly, we need to determine the correct value of $\delta_0$ (or equivalently, the peak chemical potential $\mu_F(0)$) for a given $1/k_Fa_F$, since this is what we can measure in the experiment. This can be done by integrating the Gibbs-Duhem relation using the local density approximation, and fixing the atom number.

\section{Driven scissors mode model}

Our model to explain the observed scissors mode excitation is an extension of the superfluid hydrodynamic model derived in \cite{guer99s}, which begins with the coupled hydrodynamic equations governing the bosonic superfluid density $n(\vec{r})$ and phase $S(\vec{r},t)$ evolution in the Thomas-Fermi limit,
\begin{align}
\frac{\partial n}{\partial t} + \nabla \cdot (n\vec{v}) = 0 \label{eqn:continuity}
\end{align}
\begin{align}
m\frac{\partial \vec{v}}{\partial t} + \nabla \left(V_\text{eff}(\vec{r}) + g n(\vec{r}) + \frac{m\vec{v}^2}{2} \right) = 0, \label{eqn:force}
\end{align}
where $\vec{v} = (\hbar/m) \nabla S$ is the superfluid velocity, $g = 4\pi \hbar^2 a/m$ is the Bose-Bose coupling constant, $m$ is the Bose particle mass, and $V_\text{eff}(\vec{r})$ is the effective potential felt by the boson, consisting of both the effects of the trapping potential and the interspecies mean-field due to the presence of the fermionic superfluid. For simplicity of notation, we forgo the subscript ``$B$'' on all bosonic variables, but retain that for the fermions. 

Because of the large trap frequency ratio for Li and Yb, $\omega_F/\omega_B = 8$, the oscillatory motion of Yb within the Li superfluid cannot excite scissors oscillations (i.e. no backaction) and Li adiabatically transfers any imparted momentum or angular momentum to the trap. This is corroborated by our observation that the Li position and size are unaffected during the Yb dipole oscillation, and from full numerical simulations of coupled Gross-Pitaevskii equations  for the two superfluids \cite{foot_forbes16s}.

Therefore, we can write the effective potential as $V_\text{eff}(\vec{r}) = V_T(\vec{r}+y_0(t)\hat{y}) + g_{BF}n_F(\vec{r} + \vec{d}(t))$, where $\vec{d}(t) = (x_0,y_0(t),0)$, $x_0$ is the horizontal displacement between the two cloud centers, and we center the coordinate system on the center of the Yb cloud. For the purpose of this derivation, we assume the displacement is entirely in the $x$ direction. The trapping potential is given by $V_T(x,y,z) = \frac{m}{2} \left(\omega_x^2 x^2 +\omega_y^2 y^2 +\omega_z^2 z^2  \right)$.

As done in \cite{guer99s}, we proceed by reducing Eqns. (\ref{eqn:continuity})-(\ref{eqn:force}) to extract the linear-response dynamics of the quadrupole variables $\langle xy \rangle$ and $\langle xv_y +yv_x\rangle$, where $\langle f(\vec{r})\rangle = \int d^3\vec{r} f(\vec{r})n(\vec{r})/N$. In this way, we find
\begin{align}
\frac{d \langle xy \rangle}{dt} = \langle xv_y +yv_x \rangle
\label{eqn:contreduced}
\end{align}
\begin{align}
\frac{d \langle xv_y + yv_x\rangle}{dt} = -\frac{1}{m} \left\langle x\frac{\partial V_\text{eff}}{\partial y}+y\frac{\partial V_\text{eff}}{\partial x}\right\rangle.
\label{eqn:forcereduced}
\end{align}
Reducing the coupled first-order equations to a single second-order equation, we find
\begin{align}
\frac{d^2\langle xy\rangle}{dt^2} =& -(\omega_x^2+\omega_y^2)\langle xy \rangle - \omega_y^2y_0(t)\langle x \rangle \nonumber \\
&- \frac{g_{BF}}{m}\left\langle x\frac{\partial n_F(\vec{r}+\vec{d}(t))}{\partial y} +y\frac{\partial n_F(\vec{r}+\vec{d}(t))}{\partial x} \right\rangle \label{eqn:firstode}.
\end{align}
We then take as our ansatz for the condensate density \cite{guer99s}
\begin{align}
n(\vec{r},t) = \frac{1}{g}\text{Max}\left(\mu - \frac{m}{2}\left(\omega_x^2x^2 + \omega_y^2 y^2 + \omega_z^2 z^2\right.\right.& \nonumber \\
\left.\left.+ (\omega_x^2 +\omega_y^2)\alpha(t) xy\right),0\right)&, \label{eqn:ansatz}
\end{align}
allowing us to define the tilt angle $\theta(t)$ from
\begin{align}
\tan(2\theta(t)) = \frac{(\omega_x^2+\omega_y^2)\alpha(t)}{\omega_y^2-\omega_x^2}. \label{eqn:rotationangle}
\end{align}
Since we expect the tilt of the BEC to be small, we can expand the density as follows,
\begin{align}
n(\vec{r},\theta(t)) &= n(\vec{r},0) + \theta(t)\left.\frac{\partial n}{\partial \theta}\right\vert_{\theta=0} + \mathcal{O}(\theta^2). \label{eqn:densityexpansion}
\end{align}

\begin{figure}
\centering
\includegraphics[width=1\columnwidth]{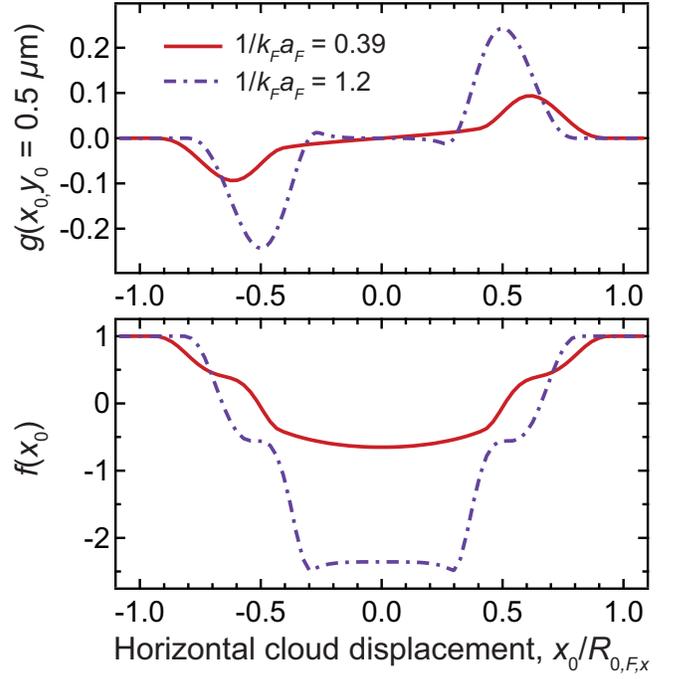}
\caption{\label{fig:scissorsfig} Analytical model for the driven scissors mode in linear response. (Upper panel) Function $g(x_0,y_0)$, which quantifies the magnitude of the scissors ``forcing'', evaluated at $1/k_Fa_F = 0.39$ (red solid line) and $1.2$ (purple dot-dashed line) for the observed dipole oscillation in-trap amplitude $y_0 = 0.5$ $\mu$m, plotted vs. horizontal cloud displacement in units of the ideal gas Fermi radius $R_{0,F,x}$. For these calculations, we assume the displacement lies along the horizontal direction with trap frequency $2\pi \times 59$ Hz. (Lower panel) The function $f(x_0)$ associated with the displacement-dependent shift of the scissors mode natural frequency.}
\end{figure}

One can show that, to first order in $\theta$, the first term on the RHS of Eqn. (\ref{eqn:firstode}) evaluates to
\begin{align}
\langle xy \rangle = -\frac{2(\omega_y^2-\omega_x^2)\theta(t)}{3m\omega_x^2\omega_y^2}\left\langle V_T(\vec{r}) \right\rangle, \label{eqn:xyavgval}
\end{align}
and the second term vanishes. Here $\left\langle V_T(\vec{r})\right\rangle = 3\mu/7$ is the trap-averaged potential energy in the BEC. For the third term on the RHS of Eqn. (\ref{eqn:firstode}), we find
\begin{align}
\frac{g_{BF}}{m}\left\langle \cdot\cdot\cdot \right\rangle = I_1(x_0,y_0(t)) + \theta(t)I_2(x_0,y_0(t)),
\end{align}
where $I_1$ and $I_2$ are integrals over the BEC distribution, and we use the Fermi superfluid density profiles discussed in the previous section. We can then expand these functions in the small oscillation amplitude $y_0(t) = y_0\cos(\omega_y't)$, giving
\begin{align}
I_1(x_0,y_0(t)) \approx I_1(x_0,0) &+ y_0\cos(\omega_y' t)\left.\frac{\partial I_1}{\partial y_0}\right\vert_{y_0=0}, \label{eqn:int1expansion}
\end{align}
and similarly for $I_2$. Note that this is the only place where the shifted dipole oscillation frequency $\omega_y'$ enters into the scissors analysis. Due to the symmetries of the integrands at $y_0=0$, we always have $I_1(x_0,0) = \partial I_2(x_0,0)/\partial y_0 = 0$. Finally, we can write the governing equation (\ref{eqn:firstode}) in terms of the tilt angle using Eqn. (\ref{eqn:xyavgval}), and recover Eqn. (2) of the main text. The functions $f(x_0)$ and $g(x_0,y_0)$ introduced in the main text are shown in Fig. \ref{fig:scissorsfig} as a function of the horizontal displacement in units of ideal gas Fermi radius, $R_{0,F,x}$. Note that, since $I_2$ has a vanishing first derivative at $y_0 = 0$, the resulting function $f$ does not depend on $y_0$ to lowest order. For these calculations, we use the higher of the two horizontal trap frequencies, which is $\omega_x = 2\pi \times 59$ Hz.

To apply our model to the scissors mode observations (Fig. 3 of the main paper) we add a heuristic damping term to the dynamics in Eqn. (\ref{eqn:firstode}). The resulting differential equation for the in-trap angle $\theta(t)$ then becomes
\begin{align}
\frac{d^2\theta}{dt^2} = -\omega_s^2\theta - \frac{\omega_s}{Q_s}\frac{d\theta}{dt} + g(x_0,y_0)\omega_x^2\cos(\omega_y't),
\end{align}
where $\omega_s^2 = \omega_y^2 + f(x_0)\omega_x^2$ is the natural frequency of the scissors mode and $Q_s$ is the quality factor. The steady state response is $\theta(t) = \theta_0\cos(\omega_{y}'t+\phi)$, where $\theta_0 \approx Q_s g(x_0, y_0) \omega_{x}^2/(\omega_y'\omega_s)$ for $\omega_y' \approx \omega_s$. Though in principle this prescription determines the phase $\phi$ relative to the drive, we do not expect the heuristic damping model to capture this aspect because of the strong dependence of $\phi$ on the driving frequency close to resonance. 


During time-of-flight, the initial rotation of the long-axis of the BEC evolves according to superfluid hydrodynamics \cite{edwa02s,modu03s}. Our ToF of $30\,$ms is much longer than the inverse horizontal trap frequency. This results in a factor of $\approx 2$ amplification of the observed amplitude with respect to the in-trap scissors amplitude \cite{modu03s}. Using the observed dipole oscillation amplitude of $y_0 = 0.5$ $\mu$m and a horizontal displacement of $x_0 = 0.6R_{0,F,x}$ (see Fig. \ref{fig:scissorsfig}), our model matches the amplitude of the scissors mode observed in ToF for $Q_s \approx 4$, and verifies that the driving force is nearly equal for $1/k_Fa_F = 0.39$ and 1.2.

\section{\label{sec:angmo} Angular momentum in the scissors mode}

In order to investigate the role of angular momentum in the scissors mode oscillation, we evaluate the expectation value of the operator $L_z = m(xv_y - yv_x)$. One can show that, to first order, the superfluid density profile in Eqn. (\ref{eqn:ansatz}) does not change in shape during the scissors oscillations. Thus no shape oscillations are excited and the superfluid flow is incompressible, or $\nabla \cdot \vec{v} = 0$. In combination with the irrotationality constraint $\nabla \times \vec{v} = 0$, this implies that $\vec{v} = \gamma(t)\nabla(xy)$, for some $\gamma(t)$ \cite{mara02s}. Using Eqns. (\ref{eqn:continuity}), (\ref{eqn:ansatz}), and (\ref{eqn:densityexpansion}) we find that $\gamma(t) = \dot{\theta}(t)(\omega_y^2-\omega_x^2)/(\omega_y^2+\omega_x^2)$ and
\begin{align}
\left\langle L_z \right\rangle(t) &= \frac{m}{N} \dot{\theta}(t)\frac{\omega_y^2-\omega_x^2}{\omega_y^2+\omega_x^2}\int d^3\vec{r} (x^2 - y^2)n(\vec{r},t) \nonumber \\
&= \frac{2}{3}\frac{(\omega_y^2-\omega_x^2)^2}{\omega_y^2\omega_x^2(\omega_y^2+\omega_x^2)}\left\langle V_T \right\rangle \dot{\theta}(t)
\end{align}
Thus, we see directly that a time-dependent in-trap angle necessitates the existence of angular momentum in the condensate.

\section{Correction to Frequency Shift Prediction for Displaced Cloud Centers}
\begin{figure}
\begin{centering}
\includegraphics[width=1\columnwidth]{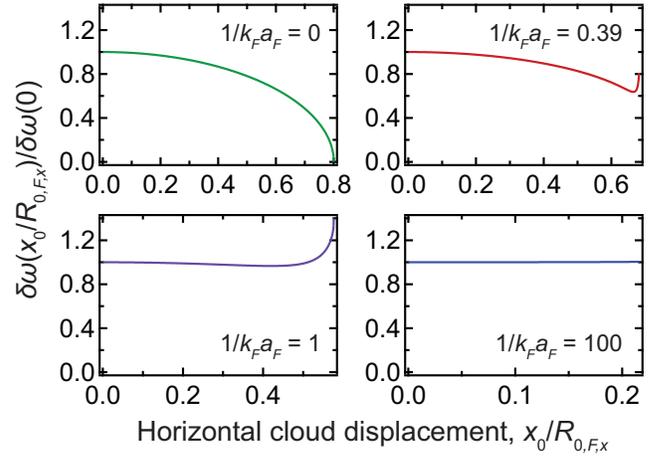}
\par\end{centering}
\caption{\label{fig:freqshiftfig} Change in the dipole frequency shift as a function of horizontal displacement of the Bose-Fermi cloud centers. Note the changing horizontal scale in each plot, since the Fermi radius $R_{F,x}$ decreases with increasing $1/k_Fa_F$. In the far BEC limit, the curvature of the Fermi gas density distribution becomes constant as expected.}
\end{figure}

For a finite displacement of the Bose and Fermi cloud centers, there is a corresponding decrease in the curvature of the mean-field interaction energy as seen by the bosonic component, resulting in a smaller value for the dipole oscillation frequency shift. We first present an analytical solution for the spatially-dependent, mean-field dipole frequency shift for a BEC with a horizontal displacement of $\vec{d} = (d_x,0,d_z)$ with respect to the center of a unitary Fermi gas.

To begin with, we have the overall Bose potential $V_B(\vec{r}-\vec{d}) = V_{T,B}(\vec{r}-\vec{d})+ V_{BF}(\vec{r})$, where $V_{T,B}(\vec{r}) = \frac{m_B}{2}(\omega_{x,B}^2x^2+\omega_{y,B}^2y^2+\omega_{z,B}^2z^2)$ is the external trapping potential for the boson, $V_{BF}(\vec{r}) = g_{BF}n_F(\vec{r})$ is the Bose-Fermi interaction energy, and we center our coordinate system on the Fermi cloud center. At $1/k_Fa_F=0$ we have 
\begin{align}
n_F(\vec{r}) = n_F\left[\mu_F(\vec{r})\right]=\gamma \mu_F(\vec{r})^{3/2}, \label{eqn:unitaryEOS}
\end{align}
where $\gamma = (2m_F/\xi\hbar^2)^{3/2}/(3\pi^2)$, $\mu_F(\vec{r}) = \mu_F(0) - V_{T,F}(\vec{r})$ (local-density approximation), $\mu_F(0) = \sqrt{\xi}E_F$, and $V_{T,F}(\vec{r}) = \frac{m_F}{2}(\omega_{x,F}^2x^2+\omega_{y,F}^2y^2+\omega_{z,F}^2z^2)$ is the external trapping potential for the fermion.

We can then get the mean-field-shifted vertical ($y$) frequency for the bosonic component at the point $\vec{d}$ from $m_B \omega_{y,B}'^2 = \left.\partial^2V_{B}/\partial y^2\right\rvert_{\vec{d}}$, giving
\begin{align}
\omega_{y,B}'^2 &= \omega_{y,B}^2 + \frac{g_{BF}}{m_B} \left.\frac{\partial^2n_F}{\partial y^2}\right\rvert_{\vec{d}}. \label{eqn:generalfreqshift}
\end{align}
Using Eqn. (\ref{eqn:unitaryEOS}) to compute $\partial^2n_F/\partial y^2$, we get
\begin{align}
\left.\frac{\delta\omega_{y,B}}{\omega_{y,B}}\right\rvert_{\vec{d}} &\approx - \frac{\beta}{2}\frac{\alpha_F}{\alpha_B}\sqrt{1-\left(\frac{d_x}{R_{x,F}}\right)^2+\left(\frac{d_z}{R_{z,F}}\right)^2},
\end{align}
where $R_{i,F} = \xi^{1/4}\sqrt{2E_F/m_F\omega_{i,F}^2}$ is the unitary Fermi cloud radius, and
\begin{align}
\beta = \frac{2(m_F+m_B)}{\pi \xi^{5/4} m_B}k_Fa_{BF}.
\end{align}
For $d_x=d_z=0$, we recover the result from \cite{ferr14s},
\begin{align}
\left.\frac{\delta\omega_{y,B}}{\omega_{y,B}}\right\rvert_0 = \frac{m_B+m_F}{\pi m_B \xi^{5/4}} \frac{\alpha_F}{\alpha_B} k_F a_{BF}.
\end{align}

For a finite horizontal displacement, we have
\begin{align}
\left.\frac{\delta\omega_{y,B}}{\omega_{y,B}}\right\rvert_{\vec{d}} = \left.\frac{\delta\omega_{y,B}}{\omega_{y,B}}\right\rvert_{0} \sqrt{1-\frac{d_x^2}{R_{x,F}^2}-\frac{d_z^2}{R_{z,F}^2}}.
\end{align}
Hence, in order to measure a frequency shift with a systematic error of 10\% less than the peak predicted value, the two cloud centers would have to be offset by 44\% of the Li cloud radius.

We can similarly perform the calculation for a zero-temperature Fermi gas with an arbitrary value of $1/k_Fa_F$, using Eqns. (\ref{eqn:zerotempprofile}) and (\ref{eqn:generalfreqshift}). The results are shown in Fig. \ref{fig:freqshiftfig}, where the horizontal displacement is scaled in units of the ideal Fermi gas radius. For the value $1/k_Fa_F = 0.39$, the change in frequency shift with displacement is much weaker than that at unitarity. Furthermore, note that in the far BEC limit, the density profile becomes $n_F(\vec{r}) \propto \mu_F(0) - V_{T,F}(\vec{r})$, and hence has constant curvature as a function of displacement.



\begin{thebibliography}{39}
\bibitem[{foo(2016{\natexlab{a}})}]{foot3_bfds}
\bibinfo{journal}{We perform a separate measurement of the stability of the
  dual superfluid at $780\,$G in the dipole oscillation trap conditions and
  confirm that the data in Fig. \ref{fig:dipoleosc} is well within a single
  $1/e$ decay time.}
\bibitem[{foo(2016{\natexlab{b}})}]{foot2_bfds}
\bibinfo{journal}{For the magnitude of $a_{BF}$, we use a weighted mean of the
  values reported in \cite{ivan11,hara11}.}
\end{thebibliography}

\end{document}